\documentclass[journal, final, onecolumn,12pt]{IEEEtran}
\makeatletter
\def\ps@headings{%
\def\@oddhead{\mbox{}\scriptsize\rightmark \hfil \thepage}%
\def\@evenhead{\scriptsize\thepage \hfil \leftmark\mbox{}}%
\def\@oddfoot{}%
\def\@evenfoot{}}
\makeatother 

\usepackage{amssymb,color}
\usepackage{cite,paralist}
\usepackage{graphicx,mathrsfs}
\usepackage{subfigure}
\usepackage{mathrsfs}
\usepackage{amsmath}
\usepackage{amsfonts}
\usepackage{subfigure}

\newtheorem{theorem}{{Theorem}}
\newtheorem{lemma}[theorem]{{Lemma}}

\newcommand{\mb}{\mathbf}
\newcommand{\qed}{\hspace*{\fill} $\Box$ \\}

\title{Network Coding Capacity of Random Wireless Networks under a SINR Model}

\author{Zhenning~Kong,~\IEEEmembership{Student Member,~IEEE,}
        Salah~A.~Aly,~\IEEEmembership{Student Member,~IEEE,}
        Emina~Soljanin,~\IEEEmembership{Senior Member,~IEEE,}
        Edmund~M.~Yeh,~\IEEEmembership{Member,~IEEE,}
        Andreas~Klappenecker~\IEEEmembership{Member,~IEEE}
\thanks{This research is supported in part by National Science Foundation (NSF) grant CNS-0626882,
Army Research Office (ARO) grant W911NF-07-1-0524.}
\thanks{Z.~Kong and Edmund~M.~Yeh are with the Department of Electrical Engineering, Yale University, New Haven, CT 06520, USA, (email: zhenning.kong@yale.edu, edmund.yeh@yale.edu).}
\thanks{S.~A.~Aly and A.~Klappenecker are with the Department of Computer Science, Texas A\&M University, College Station, TX 77843, USA, (email: salah@cs.tamu.edu, klappi@cs.tamu.edu).}
\thanks{E.~Soljanin is with Bell Laboratories, Alcatel-Lucent,
Murray Hill, NJ 07974, USA, (email: emina@lucent.com).}}

\begin{document}

\maketitle

\begin{abstract}
Previous work on network coding capacity for random wired and wireless networks have
focused on the case where the capacities of links in the network are independent. In this
paper, we consider a more realistic model, where wireless networks are modelled by random
geometric graphs with interference and noise. In this model, the capacities of links are
not independent. By employing coupling and martingale methods, we show that, under mild
conditions, the network coding capacity for random wireless networks still exhibits a
concentration behavior around the mean value of the minimum cut.
\end{abstract}

\baselineskip 20 pt

\section{Introduction}

Traditionally, information flow in networked systems was treated like fluid through
pipes, and independent information flows were processed separately. Under this
assumption, for a unicast transmission (one source node transfers information to one
destination node), the maximum transmission rate is bounded by the size of the minimum
cut between the source and the destination. This result is known as the \emph{Min-Cut
Max-Flow Theorem}, which was proved by Menger~\cite{Me27}, Ford and
Fulkerson~\cite{FoFu56} and Elias~\emph{et al.}~\cite{ElFeSh56}. However, for a multicast
transmission (one source node transfers information to multiple destination nodes), this
maximum flow rate cannot always be achieved by traditional store-and-forward routing
algorithms, even if each source-destination pair has the minimum cut with the same size.
That is because in a multicast transmission, some links in the network may be shared by
the routing paths for different source-destination pairs.

In their seminal paper~\cite{AhCaLiYe00}, Ahlswede~\emph{et al.} proposed a~\emph{network
coding} scheme, and showed that if we allow intermediate nodes to encode their received
messages and forward the coded messages to their next-hop neighbors, the maximum flow
rate can be achieved for mutilcast transmissions. In addition to the information
theoretic treatment of~\cite{AhCaLiYe00}, network coding has also been studied in an
algebraic framework developed by Koetter and M\'{e}dard in~\cite{KoMe03}, and a
combinatorial framework proposed by Fragouli and Soljanin in~\cite{FrSo06}. Code design
for network coding schemes has also attracted intense interest. In~\cite{LiYeCa03},
Li~\emph{et al.} showed that linear codes are sufficient to achieve the maximum flow rate
for a one-source multicast transmission. Koetter and M\'{e}dard, and Jaggi~\emph{et al.}
constructed linear multicast codes for network coding schemes in~\cite{KoMe03} and
in~\cite{JaSaChEfEgTo05}, respectively. The approach of constructing linear codes in a
randomized way for multicast transmissions was proposed by Ho~\emph{et al.}
in~\cite{HoKoMeEfShKa06}. For a detailed review of network coding and its applications in
many fields, e.g., wireless communication, content distribution, security, please see the
book by Fragouli and Soljanin~\cite{FrSo07}.

In most studies on network coding, network topologies are assumed to be known.
In~\cite{RaShWe03, RaShWe05}, Ramamoorthy~\emph{et al.} studied network coding capacity
for weighted random graphs and random geometric graphs. In the random graph model, each
pair of nodes are connected by a bidirectional link with probability $p<1$
independently~\cite{Bo01,JaLuRu00}. The capacity of each link is assumed to be i.i.d.
according to some probability distribution. In the random geometric graph model, two
nodes are connected to each other by a bidirectional link only when their distance is
less than or equal to a predefined positive value $r$, the characteristic
radius~\cite{Pe03}. Each link has a unit capacity. For these two types of random
networks, the authors showed that the network coding capacity is concentrated at the
(weighted) mean degree of the graph, i.e., the (weighted) mean number of neighbors of
each node. Essentially, the results reveal a concentration behavior of the size of the
minimum cut between two nodes in random graphs or random geometric graphs. Related
problems have been studied in the literature, e.g., \cite{DiPePeSe01} and references
therein. In~\cite{AlKaMeKl07}, the authors studied a generalized random geometric graph
model, where two nodes are connected by a bidirectional link with probability 1 if their
distance $d$ is less than or equal to $r_0>0$, and with probability $p<1$ if $r_0<d\leq
r_1$. They obtained similar concentration results.

The geometric models in~\cite{RaShWe03, RaShWe05, AlKaMeKl07} assume that a link exists
(possibly with a probability) between two nodes when the nodes are within each other's
transmission range. Although each link has a direction, as all links are bidirectional
(i.e., the link $(i,j)$ implies the existence of the link $(j,i)$), the model in fact
leads to an {\em undirected} graph and considerably simplifies the resulting analysis. In
addition, interference among wireless terminals was not considered in~\cite{RaShWe03,
RaShWe05, AlKaMeKl07}. Nevertheless, in wireless networks, due to noise, interference,
and the heterogeneity of transmission powers, significantly more sophisticated models for
link connectivity are needed. For instance, a widely-used model for wireless
communication channels is the Signal-to-Interference-and-Noise-Ratio (SINR)
model~\cite{Pr00, TsVi05}. In this paper, we study the capacity, i.e., the size of the
minimum cut, of random wireless networks under the SINR model.

Given that network coding capacity with noisy links is in general still an open problem,
we assume that as long as the SINR $\beta_{ij}$ of a link $(i,j)$ is greater than or
equal to a predefined threshold $\beta$, then node $i$ can transmit data at rate $R$
packets/sec to node $j$ without error. That is, links are noise-free once the SINR
condition is met. In other words, we view network coding as operating on a higher layer
in the network communication stack, and assume there is an error correcting code at the
lower layer which corrects errors on the links once the SINR threshold is met. Given this
model, each link is indeed directional, and the capacities of different links are not
independent. Nevertheless, we will show that under some mild conditions, the capacity
still has a sharp concentration when the scale of the network is large enough.

It is worthy mentioning that the capacity we investigate in this paper is different from
the one studied in~\cite{GuKu00, GrTs02, NeMo05, GaMaPrSh06, LiShMaSh06, LiGoTo07}. The
latter is referred to as throughput capacity, or transport capacity, for random wireless
networks with many-to-many transmissions. In other words, it is the maximum achievable
averaged rate at which each node in the network can transmit (simultaneously with other
nodes specified by scheduling schemes) to a randomly selected destination node. In
contrast, the network coding capacity that we study in this paper (as in~\cite{RaShWe03,
RaShWe05, AlKaMeKl07}) is the maximum rate that one source can achieve in a multicast
transmission, which is determined by the size of the minimum cut between the source and
the destinations.

This paper is organized as follows. In Section II, we describe the random wireless
network model. In Section III, we study the network coding capacity for a single source
and multiple destinations. Specifically, we investigate two cases. In the first case, all
nodes have the same transmission power, and in the second case, the transmission powers
are heterogeneous. We use different techniques for these two cases and show that the
network coding capacity has a concentration behavior in both cases. In Section IV, we
present relevant simulation results, and finally, we conclude in Section V.

\section{Random Wireless Network Model}

We use the following model for random wireless networks. Assume
\begin{itemize}
\item[(i)] $\mathcal{X}=\{{\mb X}_1, {\mb X}_2, ..., {\mb X}_n\}$ is a set of $n$ nodes
which are independently and uniformly distributed at random on the two-dimensional unit
torus, where ${\mb X}_i$ denotes the random location of node $i$, and $n$ is the total
number of nodes. \item[(ii)] Each node $i$ has a transmission power $P_i$, which follows
a probability distribution $f_P(p)$, $p\in [p_{min},p_{max}]$, where $0<p_{min}\leq
p_{max}<\infty$.
\end{itemize}
Here, the existence of a link from node $i$ to node $j$ depends on $j$'s ability to
decode the transmitted signal from $i$, which is in turn determined by the
Signal-to-Interference-and-Noise-Ratio (SINR) given by
\begin{equation}\label{eq:beta-ij}
\beta_{ij}=\frac{P_iL(d_{ij})}{N_0+\gamma \sum_{k\neq i,j} P_kL(d_{kj})},
\end{equation}
where $P_i$ is the transmission power of node $i$, $d_{ij}$ is the distance between nodes
$i$ and $j$, and $N_0$ is the power of background noise. The parameter $\gamma$ is the
inverse of the system processing gain.  It is equal to 1 in a narrow-band system and
smaller than 1 in a broadband (e.g., CDMA) system. The signal attenuation function
$L(\cdot)$ is a function of the distance $d_{ij}=||\mathbf{X}_i-\mathbf{X}_j||$, where
$\|\cdot\|$ is the Euclidean norm, and is usually given by $L(d_{ij}) =
cd_{ij}^{-\alpha}$ for some constants $c$ and $2 < \alpha < 4$.

Under the SINR model, the transmitted signal of node $i$ can be decoded at $j$ if and
only if $\beta_{ij} > \beta$, where $\beta$ is some threshold for decoding. In this case,
a link $(i,j)$ from $i$ to $j$ is said to exist. Note that even if $\beta_{ij}
> \beta$, $\beta_{ji} > \beta$ may not hold and thus the link $(j,i)$ may not exist.
Thus, the graph resulting from the SINR model is in general {\em directed.} It is clear
that link $(i,j)$ is bidirectional if and only if $\min\{\beta_{ij}, \beta_{ji}\}>\beta$.
Denote by $G(\mathcal{X},\mathcal{P},\gamma)$ the ensemble of random wireless networks
induced by the above physical model, where $\mathcal{P}=\{P_1,P_2,...,P_n\}$ represents
the set of transmission powers.

For transmission power $P$ and signal attenuation function $L(\cdot)$, we assume
\begin{itemize}
\item[(i)] $p_{min}>\beta N_0$; \item[(ii)] $\Pr(P=p_{min})>0, \Pr(P=p_{max})>0$, \item[(iii)]
$L(x)$ is continuous and strictly decreasing in $x$
\end{itemize}
for technical and practical reasons. In the remainder of this paper,
under different circumstances, we may place further constraints on
$P_i$.

The sum $\sum_{k\neq j}L(d_{kj})=\sum_{k\neq j}L(||\mathbf{X}_k-\mathbf{X}_j||)$ is a
random variable depending on the locations of all nodes in the network. Define, for all
$j=1,...,n$,
\begin{equation}\label{eq:J-j}
J(j)\triangleq\sum_{k\neq j}L(d_{kj}),
\end{equation}
\begin{equation}\label{eq:I-j}
I(j)\triangleq\sum_{k\neq j}P_kL(d_{kj}).
\end{equation}

To study the asymptotic network capacity, we will let the number of nodes $n$ go to infinity.
Since the region is fixed, this corresponds to a dense network model~\cite{Pe03, GuKu00}. Another
widely used model is the extended network model~\cite{MeRo96, DoFrTh05}, in which the number of
nodes and the area of the region both go to infinity while the ratio between them---the density of
the network, is kept constant. Both models are widely used in the literature. We will focus on the
former one in this paper.

\section{Network Coding Capacity for Single Source Transmission}

\subsection{Capacity of a Cut}

Let $C_{ij}$ be the capacity of a link $(i,j)$. We will specify the form of $C_{ij}$
later for different scenarios. Consider a single-source multiple-destination transmission
problem. Let $s$ be the source node. Suppose there are $l$ destination nodes,
$t_1,...t_l$, and $m$ relay nodes, $u_1,...u_m$. Denote the set of the destination nodes
and relay nodes by $\mathcal{T}$ and $\mathcal{R}$, respectively.  Note that $\{s\},
\mathcal{T}$ and $\mathcal{R}$ are all subsets of $\mathcal{X}$.  In this paper, we
always assume that there are no direct links between the source and its destinations.
Fig.~1 gives an example of single-source single-destination transmission.

\begin{figure}[t]
\centering
\includegraphics[width=3in]{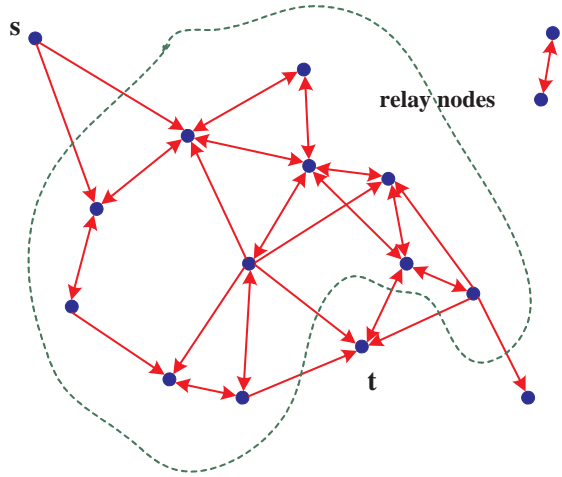}
\caption{Single-source single-destination transmission in directed SINR graphs}
\end{figure}

Let the capacity of the link from the source $s$ to each relay node $u_i$ be $C_{si}$,
$i=1,...,m$, the capacity from relay node $u_i$ to another relay node $u_j$ be $C_{ij}$,
$i\neq j, i=1,...,m, j=1,...,m$, and the capacity from each relay node $u_i$ to each
destination node $t_j$ be $C_{it_j}$, $i=1,...,m, j=1,...,l$. Unlike random geometric
graph models studied in~\cite{RaShWe03,RaShWe05, AlKaMeKl07}, the capacities in our model
are not symmetric (i.e., $C_{ij}\neq C_{ji}$) nor independent in general.

In our SINR wireless network model, there are two sources of randomness: one is the
random location of each node and the other is the random transmission power of each node.
We use $E_{X}$ and $E_{P}$ to denote the expectation operation with respect to each
probability measure, respectively.

Let $\bar{C}$ be the expected capacity of a link $(i,j)$, defined as
\begin{eqnarray}\label{C-bar}
\bar{C} & = & E_{X}E_P[C_{ij}]\nonumber \\
& = & \int_0^{\infty}C_{ij}dF_{\beta_{ij}}(\tau),
\end{eqnarray}
where $F_{\beta_{ij}}(\cdot)$ is the c.d.f. of $\beta_{ij}$, which is determined by
$f_P(\cdot)$, the distribution of $\mathcal{X}$, and the path-loss function $L(\cdot)$.

Now define an $s$-$t$-cut of size $k$ for a given source $s$ and destination $t\in
\mathcal{T}$ as a partition of the relay nodes into two subsets $V_k$ and $V_k^c$, such
that $|V_k|=k,|V_k^c|=m-k$, $V_k \cup V_k^c=\mathcal{R}$ and $V_k\cap V_k^c=\emptyset$.
An example of an $s$-$t$-cut is shown in Fig.~2. Let
\begin{equation}\label{eq:C-k}
C_k=\sum_{u_i\in V_k^c}C_{si}+\sum_{u_j\in V_k}\sum_{u_i\in V_k^c}C_{ji}+\sum_{u_j\in V_k}C_{jt},
\end{equation}
then $C_k$ is the capacity of the corresponding $s$-$t$-cut. Although $C_k$ is a sum of dependent
but identically distributed random variables, we still have
\begin{eqnarray}\label{eq:C-k-expectation}
E[C_k] & = & E_{X}E_P[C_k]\nonumber\\
& = & \sum_{u_i\in V_k^c}E_{X}E_P[C_{si}]+\sum_{u_j\in V_k}\sum_{u_i\in
V_k^c}E_{X}E_P[C_{ji}]+\sum_{u_j\in V_k}E_{X}E_P[C_{jt}]\nonumber\\
& = & [m+k(m-k)]\bar{C},
\end{eqnarray}
and consequently $E[C_k]=E[C_{m-k}]$ for $k=0,1,...,m$, and $E[C_0]\leq E[C_1]\leq \cdots \leq
E[C_{\lceil m/2 \rceil}]$.

\begin{figure}[t]
\centering
\includegraphics[width=3in]{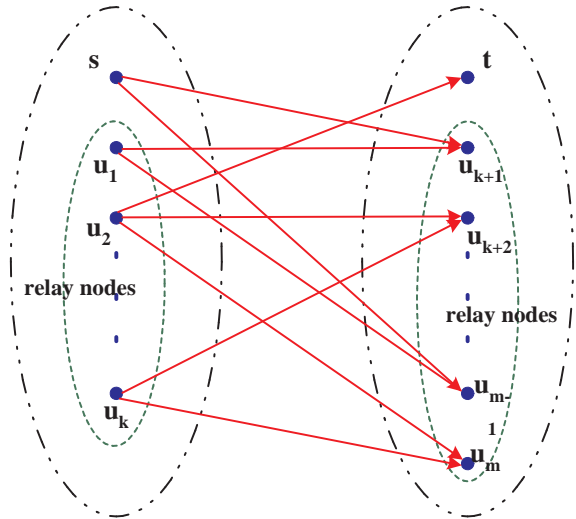}
\caption{An $s$-$t$-cut for the single-source single-destination transmission in directed SINR
graphs}
\end{figure}

To show that the capacity of any source-destination pair concentrates at some value, we
will first show that for such a source-destination pair, the capacity of any $s$-$t$-cut
of size $k$ concentrates at its mean value. Similar results were proved
in~\cite{RaShWe03, RaShWe05, AlKaMeKl07}, where the capacities of the links that
originate from the same node are i.i.d. Nevertheless, the methods used in~\cite{RaShWe03,
RaShWe05, AlKaMeKl07} do not apply here, since in the SINR model, $C_k$ is a sum of
dependent random link capacities. Instead, we employ coupling, martingale methods and
Azuma's inequality~\cite{Pe03, MoRa95} to solve the problem.

Note that when $\gamma=0$, i.e., there is no interference in the network, the capacities
$C_{si}$ for $i=k+1,...,m$, are mutually independent; so are the capacities $C_{ij}$ for
any fixed $i=1,...,k$ with $j=k+1,...,m$ or $t$. In this case, although the link
capacities are still asymmetric, $\sum_{u_i\in V_k^c}C_{si}$ and $\sum_{u_i\in V_k^c\cup
\{t\}}C_{ji}$ for $j\in V_k$ become sums of independent random variables. Thus we can
apply methods similar to those used in~\cite{RaShWe03,RaShWe05,AlKaMeKl07} to obtain the
same concentration results.

\subsection{Constant Transmission Power}

Consider the scenario when all nodes transmit with a constant power $P_0$ and denote the
corresponding model by $G(\mathcal{X},P_0,\gamma)$. In this case, the SINR of link
$(i,j)$ can be rewritten as
\begin{eqnarray}\label{eq:beta-ij-Constant-Power}
\beta_{ij} & = & \frac{L(d_{ij})}{N_0/P_0+\gamma \sum_{k\neq i,j} L(d_{kj})} \nonumber \\
& = & \frac{L(d_{ij})}{N_0/ P_0+\gamma J(j)-\gamma L(d_{ij})}.
\end{eqnarray}

Assume that when $\beta_{ij}\geq \beta$, the link $(i,j)$ has capacity $R$, i.e., node
$i$ can transmit data at rate $R$ packets/sec to node $j$ without error. Then, we can
define $C_{ij}$ as
\begin{equation}\label{eq:C-ij-R}
C_{ij}=\left\{\begin{array}{ll} R & \beta_{ij}\geq \beta,\\ 0 & \beta_{ij}<\beta.
\end{array}\right.
\end{equation}

Note that when the wireless channel is an additive Gaussian channel, the capacity of link
$(i,j)$ is~\cite{CoTh91}
\begin{equation}\label{eq:C-ij-Gaussian}
C_{ij}=\left\{\begin{array}{ll} \vspace{+.1in} \displaystyle
\frac{1}{2}\log\left(1+\beta_{ij}\right) & \beta_{ij}\geq \beta,\\ 0 & \beta_{ij}<\beta.
\end{array}\right.
\end{equation}

Our results in this subsection do not depend on any particular form of $C_{ij}$ when
$\beta_{ij}\geq \beta$. Nevertheless, since we consider the application of network
coding, it would be more appropriate to focus on \eqref{eq:C-ij-R}, rather than
\eqref{eq:C-ij-Gaussian}.

Note that $\beta_{ij}$ and thus $C_{ij}$, are determined by $L(d_{ij})$ and $J(j)$.
Because of the i.i.d. distribution of the $\mathbf{X}_i$'s, given $\mathbf{X}_j$, the
$d_{ij}$'s are independent for all $i \neq j$. Given node $j$, let
\begin{equation}\label{eq:E-L}
E[L]\triangleq E_{\mathbf{X}_i}[L(d_{ij})],
\end{equation}
then
\begin{equation}\label{eq:E-J}
E[J(j)] = E\left[\sum_{i\neq j}L(d_{ij})\right]=(n-1)E[L]\triangleq E[J].
\end{equation}

Since our model is a dense network model and the area of the region is fixed, $E[L]=E[L(d_{ij})]$
is a constant and $E[J]=(n-1)E[L]$ scales with $n$. For different $j$'s, it is clear that $J(j)$'s
are not independent. However, they have the same sharp concentration behavior in large-scale
wireless networks. This is established in the following lemma.

\vspace{+.1in}
\begin{lemma}\label{Lemma:J-j-bound}
Suppose there are $n$ nodes in the network, then
\begin{equation}\label{eq:J-j-concentration-lower}
\Pr(J(j)\leq (1-\epsilon_1)E[J])=O\left(\frac{1}{n^2}\right),
\end{equation}
and
\begin{equation}\label{eq:J-j-concentration-upper}
\Pr(J(j)\geq (1+\epsilon_1')E[J])=O\left(\frac{1}{n^2}\right),
\end{equation}
for all $j=1,2,...,n$, where $\epsilon_1=\sqrt{\frac{4\ln n}{(n-1)E[L]}}$ and $\epsilon_1'=\sqrt{\frac{6\ln n}{(n-1)E[L]}}$.
\end{lemma}
\vspace{+.1in}

\emph{Proof:}  Given any node $j$, because $J(j)=\sum_{i\neq j}L(d_{ij})$, and $L(d_{ij})$ are
i.i.d. for all $i\neq j$, by the Chernoff bound~\cite{MoRa95, AlSp00}, we have
\begin{eqnarray}\label{eq:J-j-Chernoff-lower-bound}
\!\!\!\!\!\!\!\!\Pr(J(j)\leq (1-\epsilon_1)E[J]) \!\!\!\! & \leq & \!\!\!\! \exp\left\{-\frac{E[J]\epsilon_1^2}{2}\right\}\nonumber\\
 \!\!\!\! & = &  \!\!\!\!\exp\left\{-\frac{(n-1)E[L]\epsilon_1^2}{2}\right\},
\end{eqnarray}
and
\begin{eqnarray}\label{eq:J-j-Chernoff-upper-bound}
\!\!\!\!\!\!\!\!\Pr(J(j)\geq (1+\epsilon_1')E[J]) \!\!\!\! & \leq &  \!\!\!\! \exp\left\{-\frac{E[J]\epsilon_1'^2}{3}\right\}\nonumber\\
\!\!\!\! & = & \!\!\!\! \exp\left\{-\frac{(n-1)E[L]\epsilon_1'^2}{3}\right\}.
\end{eqnarray}

Substituting $\epsilon_1=\sqrt{\frac{4\ln n}{(n-1)E[L]}}$ and $\epsilon_1'=\sqrt{\frac{6\ln
n}{(n-1)E[L]}}$ into \eqref{eq:J-j-Chernoff-lower-bound} and \eqref{eq:J-j-Chernoff-upper-bound},
we obtain \eqref{eq:J-j-concentration-lower} and \eqref{eq:J-j-concentration-upper}, respectively. \qed

Lemma \ref{Lemma:J-j-bound} shows that when the network is large, i.e., $n$ is
sufficiently large, $J(j)$ concentrates at $\gamma (n-1)E[L]=\Theta(n) E[L]$. The reason
for this is the uniform distribution of the nodes.

Now define two other types of SINR models $G'(\mathcal{X},P_0,\gamma)$ and
$G''(\mathcal{X},P_0,\gamma)$ which are coupled with $G(\mathcal{X},P_0,\gamma)$ such that they
have the same point process $\mathcal{X}$ and constant power $P_0$. Let the SINR of link $(i,j)$
in $G'(\mathcal{X},P_0,\gamma)$ and $G''(\mathcal{X},P_0,\gamma)$ be
\begin{equation}\label{eq:beta-ij-prime-1}
\beta_{ij}'=\frac{L(d_{ij})}{N_0/P_0+(1+\epsilon_1')\gamma E[J]-\gamma L(d_{ij})}
\end{equation}
and
\begin{equation}\label{eq:beta-ij-prime-prime-1}
\beta_{ij}''=\frac{L(d_{ij})}{N_0/P_0+(1-\epsilon_1)\gamma E[J]-\gamma L(d_{ij})},
\end{equation}
respectively.

Let $C_{ij}'$ and $C_{ij}''$ be the capacity of link $(i,j)$ in $G'(\mathcal{X},P_0,\gamma)$ and
$G''(\mathcal{X},P_0,\gamma)$, respectively. Since $\epsilon_1\rightarrow 0$ and
$\epsilon_1'\rightarrow 0$ as $n\rightarrow\infty$, $C_{ij}'$ and $C_{ij}''$ are asymptotically
equal to $C_{ij}$.

The following lemma establishes a concentration result for $C_k$ with constant transmission power
by coupling methods.
\vspace{+.1in}%
\begin{lemma}\label{Lemma:C-k-bound-1}
For any $0<\epsilon<1$, the capacity of an $s$-$t$-cut of size $k, k=0,1,...,m$, satisfies
\begin{equation}\label{eq:C-k-concentration-lower-1}
\Pr(C_k\leq (1-\epsilon)E[C_k']) \leq
\exp\left\{-\frac{E[C_k']\epsilon^2}{2}\right\}\left(1-O\left(\frac{1}{n}\right)\right)+
O\left(\frac{1}{n}\right)
\end{equation}
where $E[C_k']=[m+k(m-k)]\bar{C}'$ and $\bar{C}'$ is the average link capacity in
$G'(\mathcal{X},P_0,\gamma)$. Moreover,
\begin{equation}\label{eq:C-k-concentration-upper-1}
\Pr(C_k\geq (1+\epsilon)E[C_k''])  \leq
\exp\left\{-\frac{E[C_k'']\epsilon^2}{3}\right\}\left(1-O\left(\frac{1}{n}\right)\right)
+ O\left(\frac{1}{n}\right)
\end{equation}
where $E[C_k'']=[m+k(m-k)]\bar{C}''$ and $\bar{C}''$ is the average link capacity in $G''(\mathcal{X},P_0,\gamma)$.
\end{lemma}
\vspace{+.1in}

\emph{Proof:} By Lemma \ref{Lemma:J-j-bound}, for all $j$, $\{J(j)\geq (1-\epsilon_1)E[J]\}$ and
$\{J(j)\leq (1+\epsilon_1')E[J]\}$ are both increasing events.\footnote{In the context of graph
theory, an event $A$ is called increasing if $I_A(G)\leq I_A(G')$ whenever graph $G$ is a subgraph
of $G'$, where $I_A$ is the indicator function of $A$. An event $A$ is called decreasing if
$A^{c}$ is increasing. For details, please see~\cite{Pe03, AlSp00, MeRo96}.} By the FKG inequality
\cite{Pe03, AlSp00, MeRo96}, we have
\begin{eqnarray}\label{eq:FKG1}
\Pr\left(\bigcap_{j=1}^n \{J(j)\geq (1-\epsilon_1)E[J]\}\right)
&  \geq &  \prod_{j=1}^n\Pr(J(j)\geq (1-\epsilon_1)E[J])\nonumber\\
&= &  \left(1-O\left(\frac{1}{n^2}\right)\right)^n\nonumber\\
& = &  1-O\left(\frac{1}{n}\right),
\end{eqnarray}
where the first equality is due to Lemma \ref{Lemma:J-j-bound}. Similarly,
\begin{eqnarray}\label{eq:FKG2}
\Pr\left(\bigcap_{j=1}^n \{J(j)\leq (1+\epsilon_1')E[J]\}\right)
& \geq &  \prod_{j=1}^n\Pr(J(j)\leq (1+\epsilon_1')E[J])\nonumber\\
& = &  \left(1-O\left(\frac{1}{n^2}\right)\right)^n\nonumber\\
&  = & 1-O\left(\frac{1}{n}\right).
\end{eqnarray}

Inequalities \eqref{eq:FKG1} and \eqref{eq:FKG2} imply that
\begin{equation}\label{eq:C-k-C-k-prime-prime}
\Pr(C_k\leq C_k'')\geq 1-O\left(\frac{1}{n}\right),
\end{equation}
\begin{equation}\label{eq:C-k-C-k-prime}
\Pr(C_k\geq C_k')\geq 1-O\left(\frac{1}{n}\right).
\end{equation}
Since
\[
\Pr(C_k\leq (1-\epsilon)E[C_k']) \leq 1-\Pr(C_k\geq
C_k')\Pr(C_k'\geq(1-\epsilon)E[C_k']),
\]
and
\[
\Pr(C_k\geq (1+\epsilon)E[C_k'']) \leq 1-\Pr(C_k\leq
C_k'')\Pr(C_k''\leq(1+\epsilon)E[C_k'']),
\]
in order to show \eqref{eq:C-k-concentration-lower-1} and \eqref{eq:C-k-concentration-upper-1}, it
suffices to show
\begin{equation}\label{eq:C-k-prime-concentration-1}
\Pr(C_k'\leq (1-\epsilon)E[C_k'])\leq \exp\left\{-\frac{E[C_k']\epsilon^2}{2}\right\},
\end{equation}
and
\begin{equation}\label{eq:C-k-prime-prime-concentration-1}
\Pr(C_k''\geq (1+\epsilon)E[C_k''])\leq \exp\left\{-\frac{E[C_k'']\epsilon^2}{3}\right\}.
\end{equation}

In $G'(\mathcal{X},P_0,\gamma)$ and $G''(\mathcal{X},P_0,\gamma)$, the SINR of link
$(i,j)$ is given by \eqref{eq:beta-ij-prime-1} and \eqref{eq:beta-ij-prime-prime-1},
respectively. Because the $d_{ij}$'s for a given $i$ are independent, by applying the
Chernoff bounds, we obtain \eqref{eq:C-k-prime-concentration-1} and
\eqref{eq:C-k-prime-prime-concentration-1}.\qed

Since $C_{ij}'$ and $C_{ij}''$ are asymptotically equal to $C_{ij}$, $E[C_k']$ and
$E[C_k'']$ are asymptotically equal to $E[C_k]$. Consequently, Lemma
\ref{Lemma:C-k-bound-1} shows that $C_k$ concentrates at $E[C_k]$ asymptotically almost
surely.\footnote{An event is said to be asymptotic almost sure (abbreviated a.a.s.) if it
occurs with a probability converging to 1 as $n \rightarrow \infty$.}

Now, let $C_{s,t}$ be the minimum cut capacity among all $s$-$t$-cuts, i.e.,
\begin{equation}\label{eq:C-s-t} C_{s,t}=\min_{0\leq k\leq m}C_k.
\end{equation}
For one source and multiple destinations, the capacity of network coding depends on the
minimum cut between the source and the destinations~\cite{RaShWe03, RaShWe05,
AlKaMeKl07}. Therefore, for the given source node $s$ and the sets of destination nodes
$\mathcal{T}=\{t_1,...,t_l\}$ and relay nodes $\mathcal{R}=\{u_1,...,u_m\}$, define the
network coding capacity as
\begin{equation}\label{eq:C-s-T}
C_{s,\mathcal{T}}=\min_{t\in
\mathcal{T}}C_{s,t}.
\end{equation}

In the following, we show that when the number of relay nodes $m$ is sufficiently large, the
network coding capacity $C_{s,\mathcal{T}}$ concentrates at $E[C_0]=m\bar{C}$ with high
probability.
\vspace{+.1in}
\begin{theorem}\label{Theorem:Capacity-Lower-Bound-1}
When $n$ is sufficiently large, with high probability, the network coding capacity $C_{s,\mathcal{T}}$ satisfies
\begin{equation}\label{eq:Capacity-Lower-Bound-1}
\Pr(C_{s,\mathcal{T}}\geq (1-\epsilon_{\alpha}')E[C_0])= 1-O\left(\frac{l}{m^{\alpha}}\right),
\end{equation}
where $\epsilon_{\alpha}'=\sqrt{\frac{2\alpha \ln m}{E[C_0]}}$ for $\alpha> 0$ and $E[C_0]=m\bar{C}$.
\end{theorem}
\vspace{+.1in}

\emph{Proof:} Since the $C_{ij}$'s are asymptotically equal to the $C_{ij}'$'s, in order to show
\eqref{eq:Capacity-Lower-Bound-1}, it is equivalent to show
\[
\Pr(C_{s,\mathcal{T}}\geq (1-\epsilon_{\alpha}')E[C_0'])= 1-O\left(\frac{l}{m^{\alpha}}\right).
\]

Since $E[C_k']\geq E[C_0']$ for any $k=1,...,m$,
\[
\Pr(C_{s,t}\leq (1-\epsilon_{\alpha}')E[C_0'])\leq \Pr(C_{s,t}\leq
(1-\epsilon_{\alpha}')E[C_{k'}']),
\]
for any $t\in \mathcal{T}$, where $k'$ is the size of the minimum $s$-$t$-cut. By
\eqref{eq:C-k-concentration-lower-1} of Lemma \ref{Lemma:C-k-bound-1}, we have for
sufficiently large $n$,
\begin{eqnarray*}
\Pr(C_{s,t}\leq (1-\epsilon_{\alpha}')E[C_{k'}'])  & \leq &
\exp\left\{-\frac{\epsilon_{\alpha}'^2 [m+k'(m-k')]\bar{C'}}{2}\right\}\\
& \leq & \exp\left\{-\frac{\epsilon_{\alpha}'^2 m\bar{C'}}{2}\right\}.
\end{eqnarray*}
By choosing $\epsilon_{\alpha}'=\sqrt{\frac{2\alpha\ln m}{E[C_0]}}$, since $\bar{C}'$ and
$\bar{C}$ are asymptotically equal, we have for any $t\in \mathcal{T}$,
\[
\Pr(C_{s,t}\leq (1-\epsilon_{\alpha}')E[C_0']) = O\left(\frac{1}{m^\alpha}\right).
\]
By the union bound, we have
\begin{eqnarray*}
\Pr(C_{s,\mathcal{T}}\leq(1-\epsilon_{\alpha}')E[C_0']) & \leq &  \sum_{t\in \mathcal{T}}
\Pr(C_{s,t}\leq(1-\epsilon_{\alpha}')E[C_0'])\\
 & = & O\left(\frac{l}{m^{\alpha}}\right).
\end{eqnarray*} \qed

\vspace{+.1in}
\begin{theorem}\label{Theorem:Capacity-Upper-Bound-1}
When $n$ is sufficiently large, with high probability, the network coding capacity $C_{s,\mathcal{T}}$ satisfies
\begin{equation}\label{eq:Capacity-Upper-Bound-1}
\Pr(C_{s,\mathcal{T}}\leq (1+\epsilon_{\alpha}'')E[C_0])= 1-O\left(\frac{1}{m^{\alpha}}\right),
\end{equation}
where $\epsilon_{\alpha}''=\sqrt{\frac{3\alpha \ln m}{E[C_0]}}$ for $\alpha> 0$ and $E[C_0]=m\bar{C}$.
\end{theorem}
\vspace{+.1in}

\emph{Proof:} Since the $C_{ij}$'s are asymptotically equal to the $C_{ij}''$'s, in order to show
\eqref{eq:Capacity-Upper-Bound-1}, it is equivalent to show
\[
\Pr(C_{s,\mathcal{T}}\leq (1+\epsilon_{\alpha}'')E[C_0''])= 1-O\left(\frac{1}{m^{\alpha}}\right).
\]
To show this, it is sufficient to consider a particular cut for a source-destination
pair, e.g., an $s$-$t$-cut with capacity $\sum_{i=1}^m C_{si}$ separating the source $s$
from all the other nodes.
\begin{eqnarray*}
\Pr(C_{s,\mathcal{T}}\geq(1+\epsilon_{\alpha}'')E[C_0''])  & \leq & \Pr\left(C_{s,t}\geq(1+\epsilon_{\alpha}'')E[C_0'']\right)\\
& \leq &  \Pr\left(\sum_{i=1}^m
C_{si}\geq(1+\epsilon_{\alpha}'')E[C_0'']\right)\\
& = &  \Pr(C_0 \geq(1+\epsilon_{\alpha}'')E[C_0''])\\
& \leq &  \exp\left(-\frac{\epsilon_{\alpha}''^2 m\bar{C}''}{3}\right)\\
& = &  O\left(\frac{1}{m^{\alpha}}\right).
\end{eqnarray*}
where the last inequality follows from \eqref{eq:C-k-concentration-upper-1}. \qed

\subsection{Heterogeneous Transmission Powers}

In this subsection, we consider the case where the transmission power of each node is
randomly chosen rather than being constant. We continue to assume that the capacity of a
link $(i,j)$ is a constant $R$, which is independent of the SINR $\beta_{ij}$, when
$\beta_{ij}\geq \beta$. In this case, $\beta_{ij}$ can be rewritten as
\begin{eqnarray}\label{eq:beta-ij-Heterogeneous}
\beta_{ij} & = & \frac{P_iL(d_{ij})}{N_0+\gamma \sum_{k\neq i,j} P_kL(d_{kj})} \nonumber \\
& = & \frac{P_iL(d_{ij})}{N_0+\gamma I(j)-\gamma P_iL(d_{ij})}.
\end{eqnarray}

Because the $P_i$'s and $\mathbf{X}_i$'s are both i.i.d., using the same technique as
that for Lemma \ref{Lemma:J-j-bound}, we can prove the following lemma.
\vspace{+.1in}
\begin{lemma}\label{Lemma:I-j-bound}
Suppose there are $n$ nodes in the network, then
\begin{equation}\label{eq:I-j-concentration-lower}
\Pr(I(j)\leq (1-\epsilon_2)E[I])=O\left(\frac{1}{n^2}\right),
\end{equation}
and
\begin{equation}\label{eq:I-j-concentration-upper}
\Pr(I(j)\geq (1+\epsilon_2')E[I])=O\left(\frac{1}{n^2}\right),
\end{equation}
for all $j=1,2,...,n$, where $\epsilon_2=\sqrt{\frac{4\ln n}{(n-1)E[P]E[L]}}$ and $\epsilon_2'=\sqrt{\frac{6\ln n}{(n-1)E[P]E[L]}}$.
\end{lemma}

Even though we have concentration results for $I(j)$, we cannot employ the same coupling
methods as in the previous section. In $G'(\mathcal{X},P_0,\gamma)$ (or
$G''(\mathcal{X},P_0,\gamma)$) as described in Section III-B, the $C_{ij}'$'s (or
$C_{ij}''$'s) are independent for all $j\neq i$ for a given $i$. In our new case,
however, this independence does not hold because all the $C_{ij}$'s depend on the
transmission power $P_i$. To deal with this dependence, we use martingale methods and
Azuma's inequality to solve our problem.

\vspace{+.1in}
\begin{theorem}[{Azuma's Inequality~\cite{AlSp00}}]\label{Theorem:Azuma's-inequality}
Let $Z_0,Z_1,...,$ be a martingale sequence such that for each $i=1,2,...,$,
\[
|Z_i-Z_{i-1}|\leq c_i
\]
almost surely, where $c_i$ may depend on $i$. Then for all $n>0$ and any $\lambda>0$,

\begin{equation}\label{eq:Azuma's-inequality-lower}
\Pr(Z_n\geq Z_0+\lambda)\leq \exp\left\{-\frac{\lambda^2}{2\sum_{i=1}^nc_i^2}\right\},
\end{equation}
and
\begin{equation}\label{eq:Azuma's-inequality-upper}
\Pr(Z_n\leq Z_0-\lambda)\leq \exp\left\{-\frac{\lambda^2}{2\sum_{i=1}^nc_i^2}\right\}.
\end{equation}
\end{theorem}
\vspace{+.1in}

\emph{Proof:} Please see e.g.~\cite{AlSp00}.\qed

To use Azuma's inequality, we need to construct a martingale. A common approach to obtain
a martingale from a sequence of (not necessarily independent) random variables is to
construct a Doob sequence. Specifically, suppose we have a sequence of random variables
$Y_1, Y_2, ..., Y_n$, which are not necessarily independent. Let $S=\sum_{i=1}^n Y_i$ and
define a new sequence of random variables $\{Z_i:i=0,1,...,n\}$ by:
\begin{equation}\label{eq:Doob-sequence}
\left\{\begin{array}{lll}
Z_0 & = & E[S]\\
Z_i & = & E_{Y_{i+1},...,Y_{n}}[S|Y_1,...,Y_i], \quad i=1,2,...,n.
\end{array}\right.
\end{equation}
Then $\{Z_i:i=0,1,...,n\}$ is a martingale and $Z_n=S$.

If we are able to upper bound the difference $|Z_i-Z_{i-1}|$ for all $i$ by some
constant, then we can apply Azuma's inequality to obtain a bound on the tail probability.
For example, if the $Y_i$'s are independent, a simple upper bound for $|Z_i-Z_{i-1}|$ is
any upper bound on $|Y_i|$. However, if the $Y_i$'s are dependent, we need to further
understand the properties of the $Y_i$'s to bound $|Z_i-Z_{i-1}|$. We approach our
problem by following this idea and using the next corollary.

\vspace{+.1in}
\begin{lemma}\label{Lemma:Azuma's-inequality}
For $n>1$, given a sequence of random variables $Y_1, Y_2, ..., Y_n$, which are not necessarily independent, let $S=\sum_i^n Y_i$. If for any $y_i,y_i'\in D_i$, where $D_i$ is the support of $Y_i$,
\[
|E[S|Y_1,...,Y_{i-1},Y_i=y_i]-E[S|Y_1,...,Y_{i-1},Y_i=y_i']|\leq c_i,
\]
almost surely, where $c_i$ may depend on $i$, then for any $\lambda>0$,
\begin{equation}\label{eq:Corollary-inequality-lower}
\Pr(S\geq E[S]+\lambda)\leq \exp\left\{-\frac{\lambda^2}{2\sum_{i=1}^nc_i^2}\right\},
\end{equation}
and
\begin{equation}\label{eq:Corollary-inequality-upper}
\Pr(S\leq E[S]-\lambda)\leq \exp\left\{-\frac{\lambda^2}{2\sum_{i=1}^nc_i^2}\right\}.
\end{equation}
\end{lemma}
\vspace{+.1in}

\emph{Proof:} We prove this corollary for the case of discrete random variables. For continuous
random variables, the proof is similar.

Define a Doob sequence with respect to $\{Y_i:i=1,...,n\}$ as in
\eqref{eq:Doob-sequence}. To simplify notation, we will write
$E_{Y_{i+1},...,Y_{n}}[S|Y_1,...,Y_i]$ as $E[S|Y_1,...,Y_i]$ when there is no ambiguity.

By the total conditional probability theorem, we have
\begin{eqnarray*}
Z_{i-1} & = &  E[S|Y_1,...,Y_{i-1}] \\
 & = &
\sum_{y\in D_i} E[S|Y_1,...,Y_{i-1},Y_i=y]\Pr(Y_i=y|Y_1,...,Y_{i-1}),
\end{eqnarray*}
and
\begin{eqnarray*}
Z_i  & = &  E[S|Y_1,...,Y_i]\\
 & = & \sum_{y\in D_i} E[S|Y_1,...,Y_i]\Pr(Y_i=y|Y_1,...,Y_{i-1}).
\end{eqnarray*}
Therefore,
\begin{eqnarray*}
& &|Z_i-Z_{i-1}| \\
& = & |\sum_{y\in D_i} E[S|Y_1,...,Y_i]\Pr(Y_i=y|Y_1,...,Y_{i-1}) -
\sum_{y\in D_i} E[S|Y_1,...,Y_{i-1},Y_i=y]\Pr(Y_i=y|Y_1,...,Y_{i-1})\\
& \leq & \sum_{y\in D_i} |E[S|Y_1,...,Y_i]-E[S|Y_1,...,Y_{i-1},Y_i=y]|\Pr(Y_i=y|Y_1,...,Y_{i-1})\\
& \leq & \sum_{y\in D_i} c_i \Pr(Y_i=y|Y_1,...,Y_{i-1})\\
& = & c_i.
\end{eqnarray*}

Since $\{Z_i:i=0,1,...,n\}$ is a martingale with bounded difference of $|Z_i-Z_{i-1}|$, we can
apply Azuma's inequality to obtain \eqref{eq:Corollary-inequality-lower} and
\eqref{eq:Corollary-inequality-upper}. \qed

Now consider $G'(\mathcal{X},\mathcal{P},\gamma)$ and
$G''(\mathcal{X},\mathcal{P},\gamma)$ coupled with $G(\mathcal{X},\mathcal{P},\gamma)$
such that they have the same point process $\mathcal{X}$ and powers $\mathcal{P}$, where
the SINR's of link $(i,j)$ in $G'(\mathcal{X},\mathcal{P},\gamma)$ and
$G''(\mathcal{X},\mathcal{P},\gamma)$ are
\begin{equation}\label{eq:beta-ij-prime-2}
\beta_{ij}'=\frac{P_iL(d_{ij})}{N_0+(1+\epsilon_2')\gamma E[I]-\gamma P_iL(d_{ij})}
\end{equation}
and
\begin{equation}\label{eq:beta-ij-prime-prime-2}
\beta_{ij}''=\frac{P_iL(d_{ij})}{N_0+(1-\epsilon_2)\gamma E[I]-\gamma P_iL(d_{ij})},
\end{equation}
respectively.

Let $C_{ij}'$ and $C_{ij}''$ be the capacity of link $(i,j)$ in $G'(\mathcal{X},\mathcal{P},\gamma)$ and $G''(\mathcal{X},\mathcal{P},\gamma)$, respectively. Then, $C_{ij}'$ and $C_{ij}''$ are asymptotically equal to $C_{ij}$.

Assume that there exist positive solutions $r_{min}'>0$, $r_{max}'>0$, $r_{min}''>0$ and
$r_{max}''>0$ for the equations
\[
\frac{p_{min}L(r_{min}')}{N_0+\gamma (1+\epsilon_2')E[I]-\gamma p_{min}L(r_{min}')}=\beta,
\]
\[
\frac{p_{max}L(r_{max}')}{N_0+\gamma (1+\epsilon_2')E[I]-\gamma p_{max}L(r_{max}')}=\beta,
\]
\[
\frac{p_{min}L(r_{min}'')}{N_0+\gamma (1-\epsilon_2)E[I]-\gamma p_{min}L(r_{min}'')}=\beta,
\]
and
\[
\frac{p_{max}L(r_{max}'')}{N_0+\gamma (1-\epsilon_2)E[I]-\gamma p_{max}L(r_{max}'')}=\beta,
\]
respectively. That is
\begin{eqnarray*}
r_{min}' & = & L^{-1}\left(\frac{\beta}{1+\gamma\beta}\cdot\frac{N_0+\gamma(1+\epsilon_2')E[I]}{p_{min}}\right),\\
r_{max}' & = & L^{-1}\left(\frac{\beta}{1+\gamma\beta}\cdot\frac{N_0+\gamma(1+\epsilon_2')E[I]}{p_{max}}\right),\\
r_{min}'' & = & L^{-1}\left(\frac{\beta}{1+\gamma\beta}\cdot\frac{N_0+\gamma(1-\epsilon_2)E[I]}{p_{min}}\right),\\
r_{max}'' & = & L^{-1}\left(\frac{\beta}{1+\gamma\beta}\cdot\frac{N_0+\gamma(1-\epsilon_2)E[I]}{p_{max}}\right).
\end{eqnarray*}

Since $L(\cdot)$ is continuous and strictly decreasing, $r_{min}'$, $r_{max}'$, $r_{min}''$ and
$r_{max}''$ are all unique. In $G'(\mathcal{X},\mathcal{P},\gamma)$
($G''(\mathcal{X},\mathcal{P},\gamma)$), any node inside the circle centered at $\mathbf{X}_i$
with radius $r_{min}'$ ($r_{min}''$) is connected to node $i$ by a bidirectional link, while any
node outside the circle centered at $\mathbf{X}_i$ with radius $r_{max}'$ ($r_{max}''$) is not
connected to node $i$.

Let $\mathcal{A}(\mathbf{X}_i,r_{min}',r_{max}')$ and
$\mathcal{A}(\mathbf{X}_i,r_{min}'',r_{max}'')$ be the two annuli with inner radius
$r_{min}'$ and outer radius $r_{max}'$, and inner radius $r_{min}''$ and outer radius
$r_{max}''$, respectively. Denote by $N(r_{min}',r_{max}')$ and $N(r_{min}'',r_{max}'')$
the number of nodes in $\mathcal{A}(\mathbf{X}_i,r_{min}',r_{max}')$ and
$\mathcal{A}(\mathbf{X}_i,r_{min}'',r_{max}'')$, respectively. It is clear that
$N(r_{min}',r_{max}')$ and $N(r_{min}'',r_{max}'')$ are binomially distributed with mean
$n\pi(r_{max}'^2-r_{min}'^2)$ and $n\pi(r_{max}''^2-r_{min}''^2)$, respectively.

Now suppose the signal attenuation function $L(x) = cx^{-\alpha}$ for some constants $c>0$ and
$2<\alpha<4$. Then,
\begin{eqnarray*}
r_{min}' & = & \left(\frac{c(1+\gamma\beta)p_{min}}{\beta[N_0+\gamma(1+\epsilon_2')E[I]]}\right)^{1/\alpha},\\
r_{max}' & = & \left(\frac{c(1+\gamma\beta)p_{max}}{\beta[N_0+\gamma(1+\epsilon_2')E[I]]}\right)^{1/\alpha},\\
r_{min}'' & = & \left(\frac{c(1+\gamma\beta)p_{min}}{\beta[N_0+\gamma(1-\epsilon_2)E[I]]}\right)^{1/\alpha},\\
r_{max}'' & = &
\left(\frac{c(1+\gamma\beta)p_{max}}{\beta[N_0+\gamma(1-\epsilon_2)E[I]]}\right)^{1/\alpha},
\end{eqnarray*}
and
\begin{eqnarray}
r_{max}'^2-r_{min}'^2 & = &\frac{B(p_{min},p_{max})}{[N_0+\gamma(1+\epsilon_2')E[I]]^{2/\alpha}},\label{eq:r-prime-square-difference}\\
r_{max}''^2-r_{min}''^2 & = &
\frac{B(p_{min},p_{max})}{[N_0+\gamma(1-\epsilon_2)E[I]]^{2/\alpha}}\label{eq:r-prime-prime-square-difference},
\end{eqnarray}
where
\begin{equation}\label{eq:B}
B(p_{min},p_{max}) =
(p_{max}^{2/\alpha}-p_{min}^{2/\alpha})\left[\frac{c(1+\gamma\beta)}{\beta}\right]^{2/\alpha}.
\end{equation}
From \eqref{eq:r-prime-square-difference} and \eqref{eq:r-prime-prime-square-difference},
we can see that both $n\pi(r_{max}'^2-r_{min}'^2)$ and $n\pi(r_{max}''^2-r_{min}''^2)$
scale with $n$ as $\frac{B}{n^{2/\alpha-1}}$, since $E[I] = (n-1)E[P]E[I]S$ scales
linearly with $n$. Now assume that there exists a constant $\eta>0$ independent of $n$
such that
\begin{equation}\label{eq:r-max-r-min-constraint}
N(r_{min}',r_{max}') \leq \eta, \quad \mbox{and} \quad N(r_{min}'',r_{max}'')\leq \eta
\end{equation}
a.a.s. This assumption effectively puts a constraint on the transmission power. For
example, if we choose $\eta=1$, then the transmission power $P$ must scale with $n$ so
that $p_{max}=O(n^{1-\alpha/2})$. Note that $r_{min}'$ and $r_{max}'$ are asymptotically
equal to $r_{min}''$ and $r_{max}''$, respectively.

The following lemma establishes a concentration result for $C_k$ with heterogeneous transmission
power and constant capacity by coupling methods and Azuma's inequality.

\vspace{+.1in}
\begin{lemma}\label{Lemma:C-k-bound-2}
For any $0<\epsilon<1$, when $n$ is sufficiently large and \eqref{eq:r-max-r-min-constraint} is guaranteed, with high probability, the capacity of an $s$-$t$-cut of size $k, k=0,1,...,m$, satisfies
\begin{equation}\label{eq:C-k-concentration-lower-2}
\Pr(C_k\leq (1-\epsilon)E[C_k'])\leq
\exp\left\{-\frac{[m+k(m-k)]\bar{C}'^2\epsilon^2}{2\eta^2R^2}\right\},
\end{equation}
where $E[C_k']=[m+k(m-k)]\bar{C}'$ and $\bar{C}'$ is the average link capacity in
$G'(\mathcal{X},\mathcal{P},\gamma)$. Moreover,
\begin{equation}\label{eq:C-k-concentration-upper-2}
\Pr(C_k\geq (1+\epsilon)E[C_k''])\leq
\exp\left\{-\frac{[m+k(m-k)]\bar{C}''^2\epsilon^2}{2\eta^2R^2}\right\},
\end{equation}
where $E[C_k'']=[m+k(m-k)]\bar{C}''$ and $\bar{C}''$ is the average link capacity in $G''(\mathcal{X},\mathcal{P},\gamma)$.
\end{lemma}
\vspace{+.1in}

\emph{Proof:} By Lemma \ref{Lemma:I-j-bound}, for all $j$, $(1-\epsilon_2)E[I]\leq I(j)
\leq (1+\epsilon_2')E[I]$ holds a.a.s. As in the proof for Lemma \ref{Lemma:C-k-bound-1},
we have that $C_k$ is lower bounded by $C_k'$, and upper bounded by $C_k''$, with
probability at least $1-O(1/n)$. Hence, in order to show
\eqref{eq:C-k-concentration-lower-2} and \eqref{eq:C-k-concentration-upper-2}, it
suffices to show
\begin{equation}\label{eq:C-k-prime-concentration-2}
\Pr(C_k'\leq (1-\epsilon)E[C_k'])\leq
\exp\left\{-\frac{[m+k(m-k)]\bar{C}'^2\epsilon^2}{2\eta^2R^2}\right\}
\end{equation}
and
\begin{equation}\label{eq:C-k-prime-prime-concentration-2}
\Pr(C_k''\geq (1+\epsilon)E[C_k''])\leq
\exp\left\{-\frac{[m+k(m-k)]\bar{C}''^2\epsilon^2}{2\eta^2R^2}\right\}.
\end{equation}

To show \eqref{eq:C-k-prime-concentration-2}, we use martingale methods. Let
$Y_1=C_{s(k+1)}', Y_2=C_{s(k+2)}',...,Y_{m-k}=C_{sm}'$, $Y_{m-k+1}=C_{1(k+1)}'$,
$Y_{m-k+2}=C_{1(k+2)}',...,Y_{m-k+k(m-k)}=C_{km}'$, and $Y_{m-k+k(m-k)+1}=C_{1t}'$,
$Y_{m-k+k(m-k)+2}=C_{2t}',...,Y_{m-k+k(m-k)+k}=C_{kt}'$. Define a Doob sequence $\{Z_i:
i=0,...,m+k(m-k)\}$ with respect to $\{Y_i,i=1,2,...,m+k(m-k)\}$ as
\[\left\{
\begin{array}{lll}
Z_0 & = & E[C_k']\\
Z_i & = & E[C_k'|Y_1,...,Y_i], \quad i=1,2,...,m+k(m-k).
\end{array}\right.
\]
Then $\{Z_i: i=0,...,m+k(m-k)\}$ is a martingale and $Z_{m+k(m-k)}=C_k'$.

Since when $i\neq u$ and $j\neq v$, $C_{ij}'$ is independent of $C_{uv}'$, dependence
exists only among $C_{ij}'$'s, $j\neq i$, for a given $i$. However, the distances
$d_{ij}$ are independent for all $j\neq i$ for given $i$. Hence the difference between
$E[C_k'|Y_1,...,Y_{l-1},Y_l=y_l]$ and $E[C_k'|Y_1,...,Y_{l-1},Y_l=y_l']$ depends only on
those $Y_h$'s which are incident on the same source node as $Y_l$. When $d_{ij}\leq
r_{min}'$, $C_{ij}'=R$, and when $d_{ij}> r_{max}'$, $C_{ij}'=0$. Moreover, the number of
nodes within the annulus $\mathcal{A}(\mathbf{X}_i,r_{min}',r_{max}')$ is upper bounded
by the constant $\eta$ a.a.s. Therefore,
\[
\begin{array}{rr}
|E[C_k'|Y_1,...,Y_{l-1},Y_l=y_l]-E[C_k'|Y_1,...,Y_{l-1},Y_l=y_l']|\\ \leq \eta R
\end{array}
\] a.a.s., where $y_l$ and $y_l'$ are either 0 or $R$.

Applying the result of Lemma \ref{Lemma:Azuma's-inequality}, we have
\eqref{eq:C-k-prime-concentration-2}. In the same manner, we can show that
\eqref{eq:C-k-prime-prime-concentration-2} holds. \qed

In the following, we show that when the number of relay nodes $m$ is sufficiently large,
the network coding capacity $C_{s,\mathcal{T}}$ concentrates at $E[C_0]=m\bar{C}$ with
high probability. The proofs are based on Lemma \ref{Lemma:C-k-bound-2} and very similar
to those for Theorem \ref{Theorem:Capacity-Lower-Bound-1} and Theorem
\ref{Theorem:Capacity-Upper-Bound-1}, and provided here for completeness.

\vspace{+.1in}
\begin{theorem}\label{Theorem:Capacity-Lower-Bound-2}
When $n$ is sufficiently large, with high probability, the network coding capacity $C_{s,\mathcal{T}}$ satisfies
\begin{equation}\label{eq:Capacity-Lower-Bound-2}
\Pr(C_{s,\mathcal{T}}\geq (1-\epsilon_{\alpha})E[C_0])= 1-O\left(\frac{l}{m^{\alpha}}\right),
\end{equation}
where $\epsilon_{\alpha}=\frac{\eta R}{E[C_0]}\sqrt{2\alpha m\ln m}$ for $\alpha> 0$ and $E[C_0]=
m\bar{C}$.
\end{theorem}
\vspace{+.1in}

\emph{Proof:} Since the $C_{ij}$'s are asymptotically equal to the $C_{ij}'$'s, in order
to show \eqref{eq:Capacity-Lower-Bound-2}, it is equivalent to show
\[
\Pr(C_{s,\mathcal{T}}\geq (1-\epsilon_{\alpha})E[C_0'])=
1-O\left(\frac{l}{m^{\alpha}}\right).
\]

Since $E[C_k']\geq E[C_0']$ for any $k=1,...,m$,
\[
\Pr(C_{s,t}\leq (1-\epsilon_{\alpha})E[C_0'])\leq \Pr(C_{s,t}\leq
(1-\epsilon_{\alpha})E[C_{k'}']),
\]
for any $t\in \mathcal{T}$, where $k'$ is the size of the minimum $s$-$t$-cut. By
\eqref{eq:C-k-concentration-lower-2} of Lemma \ref{Lemma:C-k-bound-2}, we have
\begin{eqnarray*}
\Pr(C_{s,t}\leq (1-\epsilon_{\alpha})E[C_{k'}'])  & \leq &
\exp\left\{-\frac{\epsilon_{\alpha}^2 [m+k'(m-k')]\bar{C'}^2}{2(\eta+1)^2R^2}\right\}\\
& \leq & \exp\left\{-\frac{\epsilon_{\alpha}^2 m\bar{C'}^2}{2(\eta+1)^2R^2}\right\}.
\end{eqnarray*}
By choosing $\epsilon_{\alpha}=\frac{(\eta+1)R}{E[C_0]}\sqrt{2\alpha m \ln m}$ for
$\alpha> 0$, since $\bar{C}'$ and $\bar{C}$ are asymptotically equal, for any $t\in
\mathcal{T}$,
\[
\Pr(C_{s,t}\leq (1-\epsilon_{\alpha})E[C_0']) = O\left(\frac{1}{m^\alpha}\right).
\]
By the union bound, we have
\begin{eqnarray*}
\Pr(C_{s,\mathcal{T}}\leq(1-\epsilon_{\alpha})E[C_0'])  & \leq & \sum_{t\in \mathcal{T}}
\Pr(C_{s,t}\leq(1-\epsilon_{\alpha})E[C_0'])\\
& = &  O\left(\frac{l}{m^{\alpha}}\right).
\end{eqnarray*}\qed

\vspace{+.1in}
\begin{theorem}\label{Theorem:Capacity-Upper-Bound-2}
When $n$ is sufficiently large, with high probability, the network coding capacity $C_{s,\mathcal{T}}$ satisfies
\begin{equation}\label{eq:Capacity-Upper-Bound-2}
\Pr(C_{s,\mathcal{T}}\leq (1+\epsilon_{\alpha})E[C_0])= 1-O\left(\frac{1}{m^{\alpha}}\right),
\end{equation}
where $\epsilon_{\alpha}=\frac{\eta R}{E[C_0]}\sqrt{2\alpha m \ln m}$ for $\alpha> 0$ and
$E[C_0]=m\bar{C}$.
\end{theorem}
\vspace{+.1in}

\emph{Proof:} Since the $C_{ij}$'s are asymptotically equal to $C_{ij}''$'s, in order to
show \eqref{eq:Capacity-Upper-Bound-2}, it is equivalent to show
\[
\Pr(C_{s,\mathcal{T}}\leq (1+\epsilon_{\alpha})E[C_0''])=
1-O\left(\frac{1}{m^{\alpha}}\right).
\]
To show this, it is sufficient to consider a particular cut for a pair of the source and
one destination, for instance, an $s$-$t$-cut separating the source $s$ from all the
other nodes.
\begin{eqnarray*}
\Pr(C_{s,\mathcal{T}}\geq(1+\epsilon_{\alpha})E[C_0'']) & \leq &  \Pr\left(C_{s,t}\geq(1+\epsilon_{\alpha})E[C_0'']\right)\\
& \leq & \Pr\left(\sum_{i=1}^m
C_{si}\geq(1+\epsilon_{\alpha})E[C_0'']\right)\\
& = &  \Pr(C_0 \geq(1+\epsilon_{\alpha})E[C_0''])\\
& \leq &  \exp\left\{-\frac{\epsilon_{\alpha}^2 m\bar{C}''^2}{2(\eta+1)^2R^2}\right\}\\
& = & O\left(\frac{1}{m^{\alpha}}\right).
\end{eqnarray*}
where the last inequality follows from \eqref{eq:C-k-concentration-upper-2} of Lemma
\ref{Lemma:C-k-bound-2}.\qed

\begin{figure}[t!]
\centerline{ \subfigure[Interference $J(j)$]{
\includegraphics[width=2.5in]{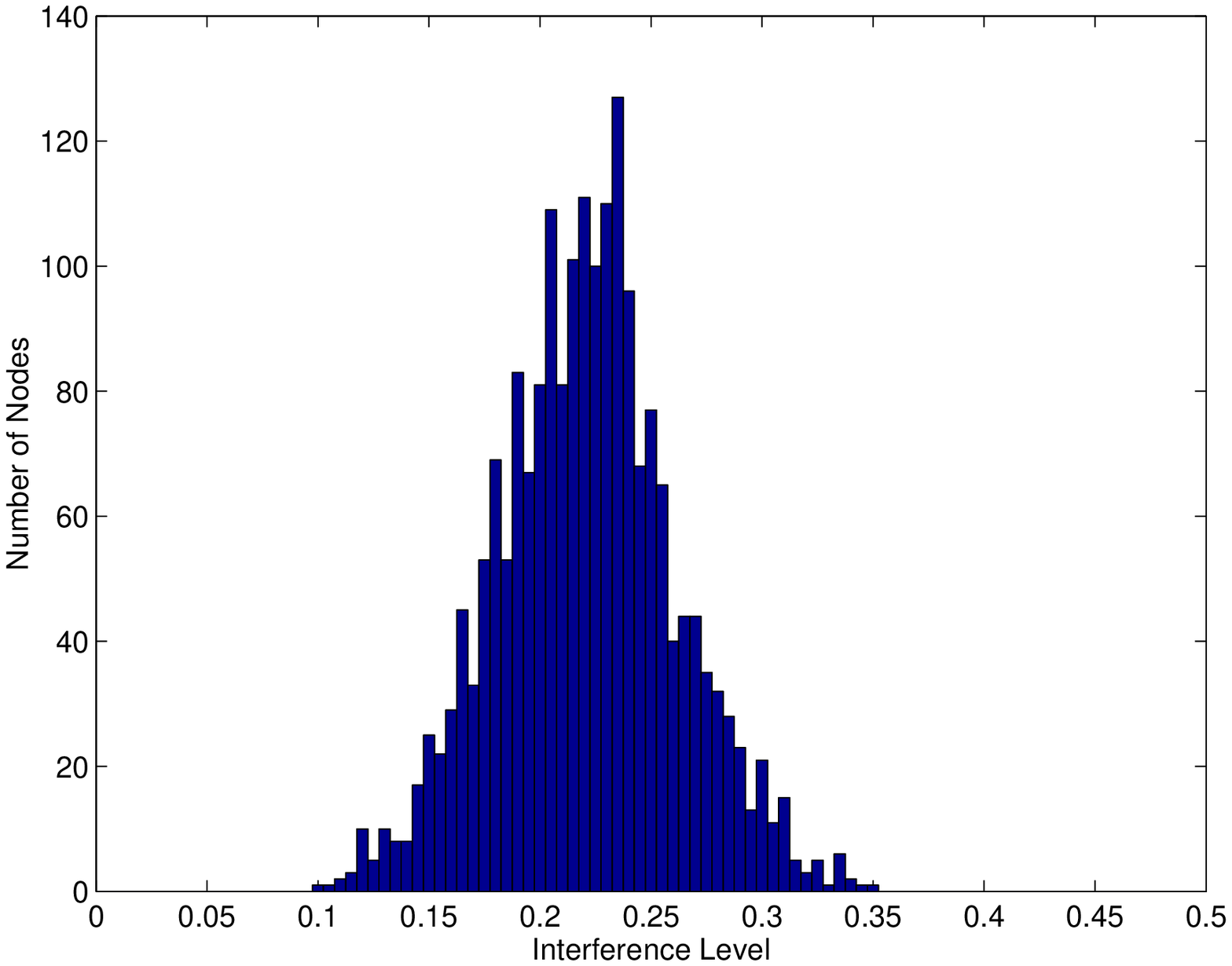}}\hfil
\subfigure[Cut capacity $C_k$]{
\includegraphics[width=2.5in]{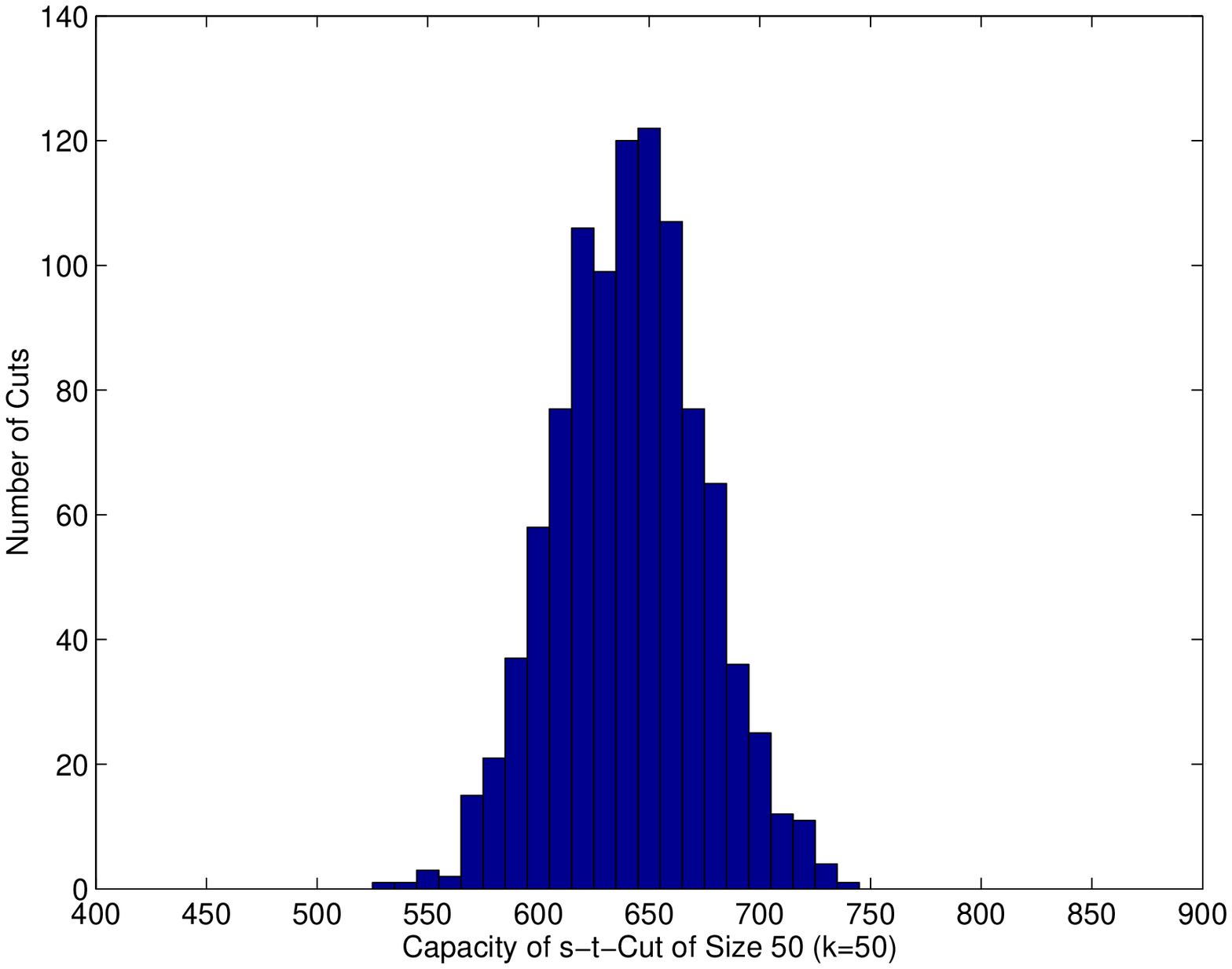}}}
\scriptsize \caption{Interference at each node, and capacity of random $s$-$t$-cut of size $k=50$
in $G(\mathcal{X},P_0,\gamma)$}
\end{figure}

\section{Simulation Studies}

In this section, we present some simulation results on the SINR model and the network
coding capacity. Fig.~3 and Fig.~4 show simulation results on interference and cut
capacity in $G(\mathcal{X},P_0,\gamma)$, where $n=2000$, $L(x)=\frac{10^{-3}}{64}x^{-3}$,
$N_0 = 0.02$, $\beta = 0.2$ and $\gamma = 0.02$, and every node transmits with constant
power $P_0=0.01$. Fig.~5 and Fig.~6 show simulation results on the interference and the
cut capacity in $G(\mathcal{X},\mathcal{P},\gamma)$, where $n=2000$,
$L(x)=\frac{10^{-3}}{64}x^{-3}$, $N_0 = 0.02$, $\beta = 0.2$ and $\gamma = 0.02$, and
every node transmits with power $P$ uniformly randomly distributed over $[0.01,0.02]$.
The results confirm the concentration behavior of the interference and the cut capacity.

\begin{figure}[t!]
\centerline{ \subfigure[Interference $J(j)$]{
\includegraphics[width=2.5in]{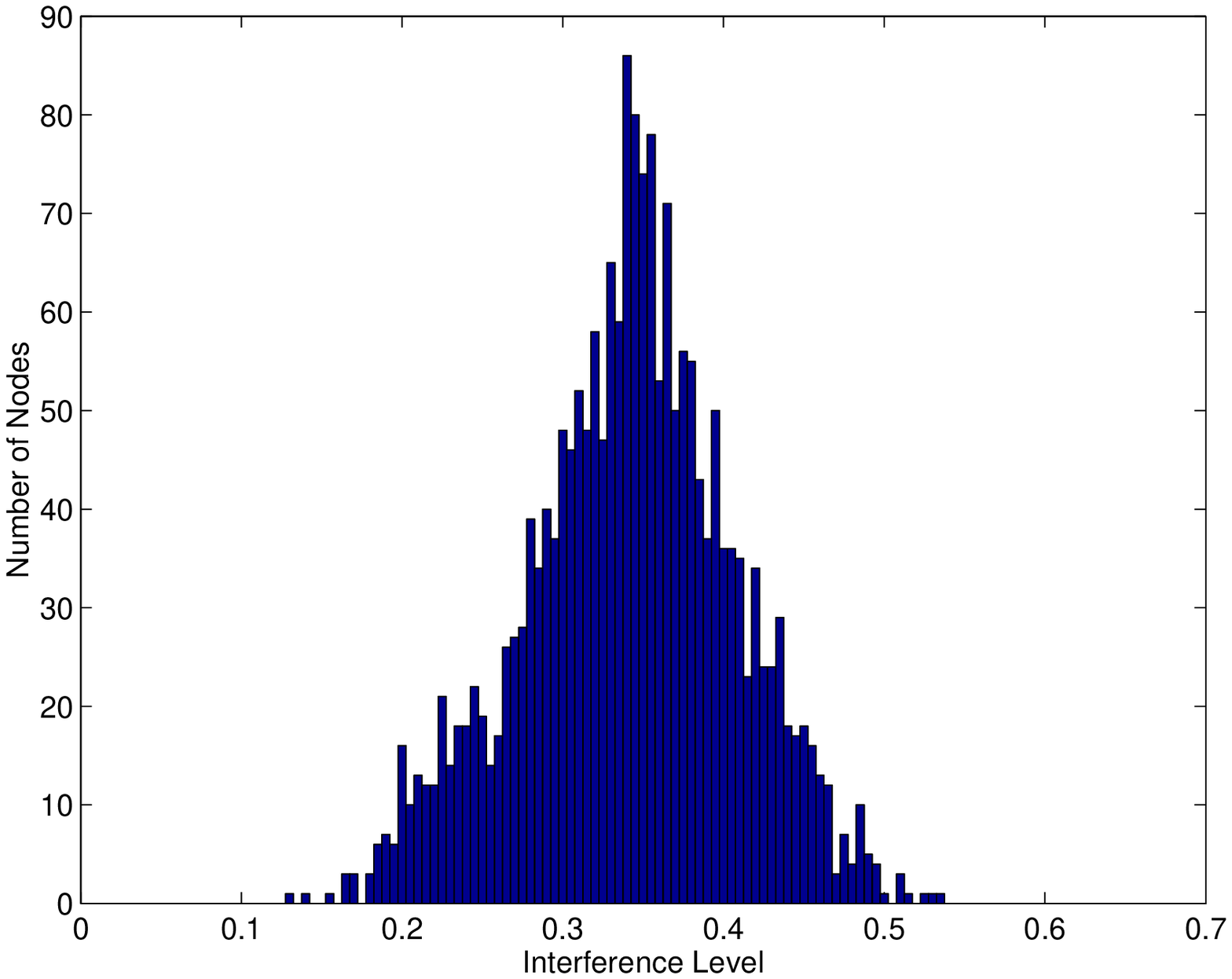}}\hfil
\subfigure[Cut capacity $C_k$]{
\includegraphics[width=2.5in]{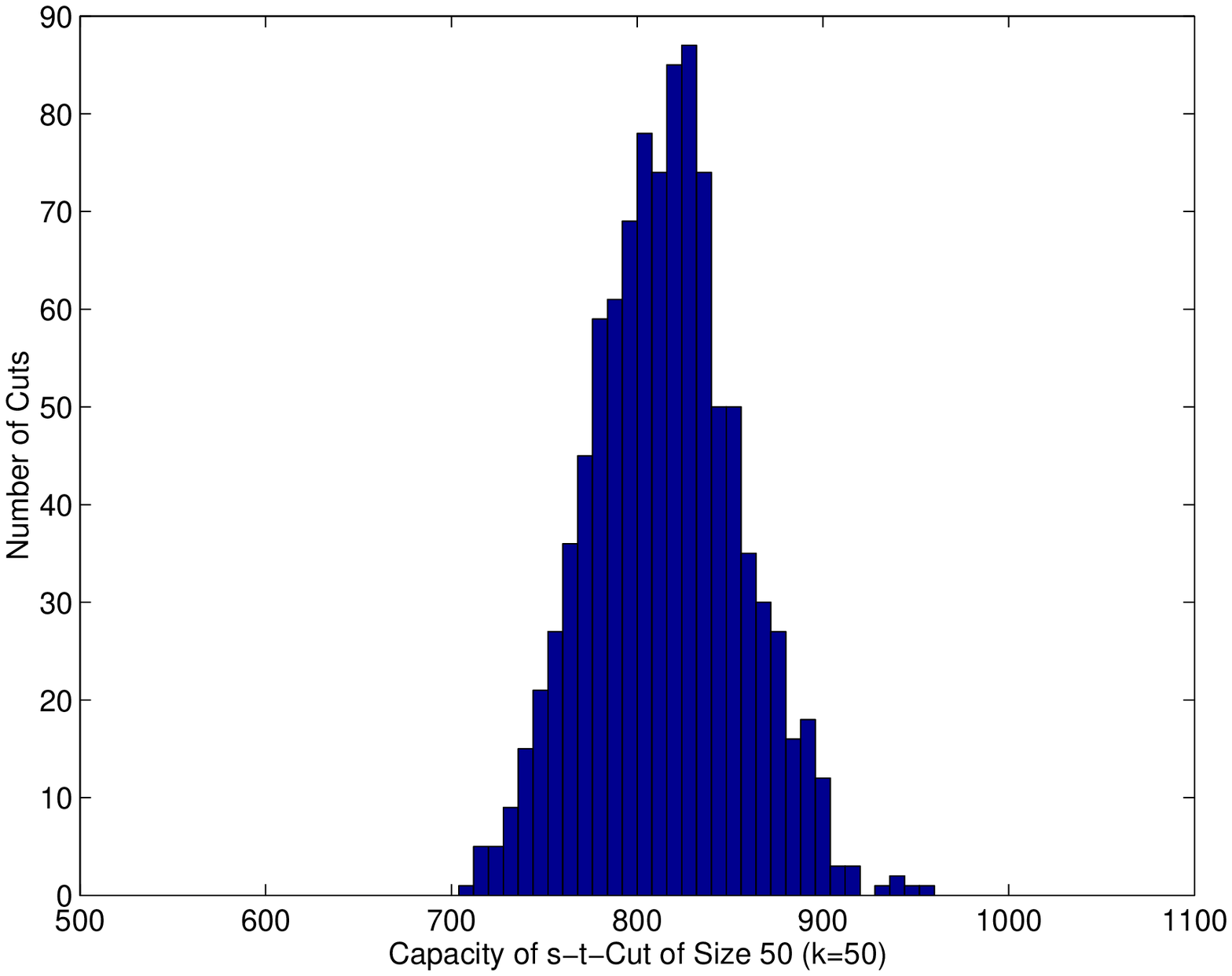}}}
\scriptsize \caption{Interference at each node, and capacity of random $s$-$t$-cut of size $k=50$
in $G(\mathcal{X},\mathcal{P},\gamma)$}
\end{figure}

\section{Conclusions}

In this paper, we studied the network coding capacity for random wireless networks under
a SINR model, where the network is modelled by the graph
$G(\mathcal{X},\mathcal{P},\gamma)$. Previous work on the network coding capacity for
random wired/wireless networks are based on the assumption that the capacities of links
are independent. In the SINR model, however, the capacities of links are not independent
due to noise and interference. We investigated two scenarios. In the first case
$G(\mathcal{X},P_0,\gamma)$, we assumed all nodes transmit with a constant power $P_0$.
To study the network coding capacity in $G(\mathcal{X},P_0,\gamma)$, we coupled
$G(\mathcal{X},P_0,\gamma)$ with two other models $G'(\mathcal{X},P_0,\gamma)$ and
$G''(\mathcal{X},P_0,\gamma)$, which have the same point process $\mathcal{X}$ and
constant power $P_0$, but different thresholds. We showed that the network coding
capacity for $G(\mathcal{X},P_0,\gamma)$ is upper and lower bounded by those for
$G'(\mathcal{X},P_0,\gamma)$ and $G''(\mathcal{X},P_0,\gamma)$. By proving that the
network coding capacities for $G'(\mathcal{X},P_0,\gamma)$ and
$G''(\mathcal{X},P_0,\gamma)$ concentrate on the same value asymptotically, we showed
that when the size of the network is sufficiently large, the network coding capacity for
$G(\mathcal{X},P_0,\gamma)$ exhibits a concentration behavior around the mean value of
the minimum cut. In the second case $G(\mathcal{X},\mathcal{P},\gamma)$, we assumed each
node transmits with a random power drawn from some distribution. Since coupling methods
could not be applied in this general case, we used martingale techniques to deal with
dependence between link capacities, and showed that under some mild conditions, the
network coding capacity also exhibits a concentration behavior. The results obtained are
important for understanding network coding performance in random wireless networks under
the SINR model. In addition, the methods used in this paper provide useful techniques for
studying properties of random wireless networks under the SINR model.

\bibliographystyle{ieeetran}

\end{document}